\documentclass[reprint,
superscriptaddress,
 amsmath,amssymb,
 aps, 
prl,
]{revtex4-2}

\pdfoutput=1

\usepackage{graphicx}
\usepackage{dcolumn}
\usepackage{bm}
\usepackage{mathtools}
\usepackage{graphicx}
\usepackage{float}
\usepackage[verbose]{placeins}
\usepackage{cleveref}
\usepackage{tikz}
\usetikzlibrary{positioning}
\usetikzlibrary{shapes.geometric}

\begin{document}
\title{Strong enhancements to superconducting properties of 1D systems from metallic reservoirs}

\author{J. E. Ebot}
\affiliation{SUPA, Institute of Photonics and Quantum Sciences, Heriot-Watt University,
Edinburgh EH14 4AS, United Kingdom}

\author{Sam Mardazad}
\affiliation{SUPA, Institute of Photonics and Quantum Sciences, Heriot-Watt University,
Edinburgh EH14 4AS, United Kingdom}

\author{Lorenzo Pizzino}
\affiliation{DQMP, University of Geneva, 24 Quai Ernest-Ansermet, 1211 Geneva, Switzerland}

\author{Johannes S. Hofmann}
\affiliation{Department of Condensed Matter Physics, Weizmann Institute of Science, Rehovot 76100, Israel}
\affiliation{Max-Planck-Institut f\"ur Physik komplexer Systeme,N\"othnitzer Strasse 38, 01187 Dresden, Germany}

\author{Thierry Giamarchi}
\affiliation{DQMP, University of Geneva, 24 Quai Ernest-Ansermet, 1211 Geneva, Switzerland}

\author{Adrian Kantian}
\affiliation{SUPA, Institute of Photonics and Quantum Sciences, Heriot-Watt University,
Edinburgh EH14 4AS, United Kingdom}

\date{\today}

\begin{abstract}
Using a 1D bilayer system comprised of pairing and metallic layers, the present work proves the striking power of reservoir-mediated boosting of superconductivity.
Employing many-body numerics on large systems at zero and finite temperature, we unravel the complex processes by which the tuning of the metal parameters can impact the effective pairing strength as well as the long-range pair-pair-coupling mediated by the metal.
It is these two processes that in turn can strongly enhance superconducting susceptibility and thermal superconducting correlation length over those of the isolated system.
We show that in this way, even a 1D system can come very close to achieving superconducting long-range order.
\end{abstract}

\maketitle

In the quest to engineer superconducting (SC) devices far beyond their current temperature limitations, one approach aims to boost the macroscopic phase coherence of electron pairs through the use of a metallic reservoir, while not degrading the strength of electron pair-binding too much at the same time.
The field arose from Kivelson's and Emery's work~\cite{Emery1995} on superconductivity emerging out of balancing two opposed goals: maximising pair-phase coherence, which requires a low effective mass of the pairing mode, and maximising pair-binding energy, which increases the mass of the pairing mode on a lattice. 
Theoretical and experimental work has aimed to show how electrons in 1D and 2D with intrinsic pairing could attempt to satisfy both goals at once, via coupling to a metallic reservoir~\cite{Kivelson2002,Erez2008,Gideon2012,Zujev2014,Dee2022,Zhang2025}.
For 2D systems, experiments on LSCO- and especially on FeSe-monolayers atop reservoirs of itinerant electrons show striking enhancements to SC $T_c$ over the isolated monolayer, but neither their cause nor their parametric dependence on reservoir properties is known~\cite{Yuli2008,Huang2017}.
To resolve these crucial questions and guide future experiments, theory studies simplified models, of a layer with an explicit pairing term (the {\rm P}-layer) in contact with a metallic one, modeled by free electrons (the {\rm M}-layer).
Yet, no enhancement above the optimized $T_c$ of the isolated 2D {\rm P}-layer could be demonstrated so far~\cite{Erez2008,Gideon2012,Zujev2014,Dee2022}, though in the strong-pairing limit an improved $T_c$ relative to the isolated baseline may have been described very recently~\cite{Zhang2025}.
However the limited linear dimensions amenable to reliable numerics in 2D greatly constrain the possible parameter ranges for both {\rm P}- and {\rm M}-layer, preventing the study of those regimes most promising for boosting SC properties.

Moving to a 1D geometry allows to sidestep these constraints due to the available analytical and numerical control and understanding.
In particular these regimes were first explored analytically for a 1D {\rm P}-layer~\cite{Lobos2009}.
That work predicted that the metallic reservoir indeed strongly boosts the stiffness of the pair-phase within the (1D) {\rm P}-layer when it can mediate long-range coupling (low-exponent algebraic decay), even sufficient to potentially attain true SC ordering at zero temperature.
Yet, non-perturbative many-body numerics are required to validate these analytical predictions, and to treat the back-action of the {\rm P}-layer onto the metal, a broader range of parameters, as well as the effects of finite temperature.
A crucial advantage of the 1D setting is that numerics can treat systems of large linear extent, where any metal-mediated long-range coupling is expected to show the most pronounced enhancement to SC properties.

This work uses strong-coupling numerics to conclusively demonstrate the strong enhancement of superconductivity
in 1D hybrid systems (c.f.~\cref{pairing1}a).
Joining {\rm P}- and {\rm M}-layer can drive SC susceptibilities far above those for the isolated {\rm P}-layer, even coming close to long-range order in a 1D system at both zero and finite temperatures.
The {\rm P}-layer exhibits a strong and superlinear enhancement of the thermal SC correlation length, another key difference to the isolated {\rm P}-layer.
Our work also reveals a powerful new effect: 
that by tuning either the nominal Fermi wave vectors of the two layers or the properties of the metal (or both), the effective pairing strength inside the P-layer, the long-range coupling mediated by the metal, and finally the SC susceptibility of the {\rm P}-layer can be tuned and optimised.

As shown in~\cref{pairing1}a, we study a 1D attractive Hubbard model coupled to a 1D metallic bath, both with $L$ sites ($L$ even), with Hamiltonian
\begin{eqnarray}
    \hat{H} &=& \sum_{i,\sigma,\lambda}^{L}-t_\lambda (\hat{c}^{\dagger}_{i,\sigma,\lambda} \hat{c}_{i+1,\sigma,\lambda} + {\rm h.c.}) - \mu_\lambda \hat{n}_{i,\sigma,\lambda}    \nonumber
    \\ &-&  t_{\perp} \sum_{i, \sigma}^{L}( \hat{c}^{\dagger}_{i,\sigma,{\rm p}} \hat{c}_{i,\sigma,{\rm m}} + {\rm h.c.}) -U\sum_{i}^{L} \hat{n}_{i,\uparrow,{\rm p}}\hat{n}_{i,\downarrow,{\rm p}},  
\end{eqnarray}
where, $\lambda ={\rm p},{\rm m}$ indexes the {\rm P}- and {\rm M}-layer respectively, and $\hat{c}_{i,\sigma,\lambda}$ is the fermionic annihilator for an electron at site $i$, of spin ${\sigma=\pm 1/2}$, in layer $\lambda$, and ${\hat{n}_{i,\sigma,\lambda}=\hat{c}^\dagger_{i,\sigma,\lambda}\hat{c}_{i,\sigma,\lambda}}$. 
The layers have separate chemical potentials $\mu_\lambda$ and tunneling $t_\lambda$, $U>0$ parametrizes the strength of pairing in the isolated {\rm P}-layer, and $t_\perp$ is the tunneling amplitude between the layers.
We study this hybrid system when inter-layer coupling $t_\perp$ is smaller than the paring gap 

\begin{figure}[htp]
    \centering
    \begin{flushleft}
       {\large(a)}
    \end{flushleft}
     \vspace{-1.07cm}
    \hspace*{0.2cm}
\begin{tikzpicture}
    [
	roundnode/.style={rectangle, draw=red!60, fill=red!5, very thick, minimum size=3mm},
	rectanglenode/.style={rectangle, draw=blue!60, fill=blue!60, very thick, minimum size=3mm,text width=2cm,align=center},
	squarednode/.style={rectangle, draw=green!60, fill=green!10, ultra thin, minimum size=3mm,text width=5cm,align=center},
    circlenode/.style={circle, draw=black!100, fill=black!100, ultra thick, minimum size=1mm,text width=0.1mm,align=center,scale=0.5},
	]
    \node[circlenode]      (sc1)  {};
    \node[circlenode]      (sc2)  [left=of sc1] {};
    \node[circlenode]      (sc3)  [left=of sc2] {};
    \node[circlenode]      (sc4)  [left=of sc3] {};
    \node[circlenode]      (sc5)  [left=of sc4] {};
    \node[circlenode]      (sc6)  [left=of sc5] {};
    %
    \node[circlenode]      (m1)  [below=of sc1] {};
    \node[circlenode]      (m2)  [left=of m1] {};
    \node[circlenode]      (m3)  [left=of m2] {};
    \node[circlenode]      (m4)  [left=of m3] {};
    \node[circlenode]      (m5)  [left=of m4] {};
    \node[circlenode]      (m6)  [left=of m5] {};
    \draw[-,ultra thick] (sc1.west)  to node[anchor=south]{\large{$t_{\rm p}$} } (sc6.east) ;
    \draw[-,ultra thick] (sc1.west)  to node[anchor=south]{\hspace*{-1.2cm}\large{$U$}} (sc2.east) ;
    \draw[-,ultra thick] (m1.west)  to node[anchor=south]{\large{$t_m$} } (m6.east) ;
    \draw[dashed,ultra thick] (sc1.south)  to  (m1.north) ;
    \draw[dashed,ultra thick] (sc2.south)  to  (m2.north) ;
    \draw[dashed,ultra thick] (sc3.south)  to  (m3.north) ;
    \draw[dashed,ultra thick] (sc4.south)  to  (m4.north) ;
    \draw[dashed,ultra thick] (sc5.south)  to  (m5.north) ;
    \draw[dashed,ultra thick] (sc6.south)  to node[anchor=east]{\large{$t_\perp $} } (m6.north) ;
    \draw[-,ultra thick] (sc1.east) -- node[anchor=west]{\large{$\mu_{\rm p} $} } (0.5,0);
    \draw[-,ultra thick] (m1.east) -- node[anchor=west]{\large{$\mu _m $} } (0.5,-1.18);
    \draw[-,ultra thick]  (sc6.west) --  (-6.40,0.0);
    \draw[-,ultra thick] (m6.west) -- (-6.4,-1.18);
\end{tikzpicture}
\includegraphics*[width=0.95\linewidth,trim= 0 30 0 85]{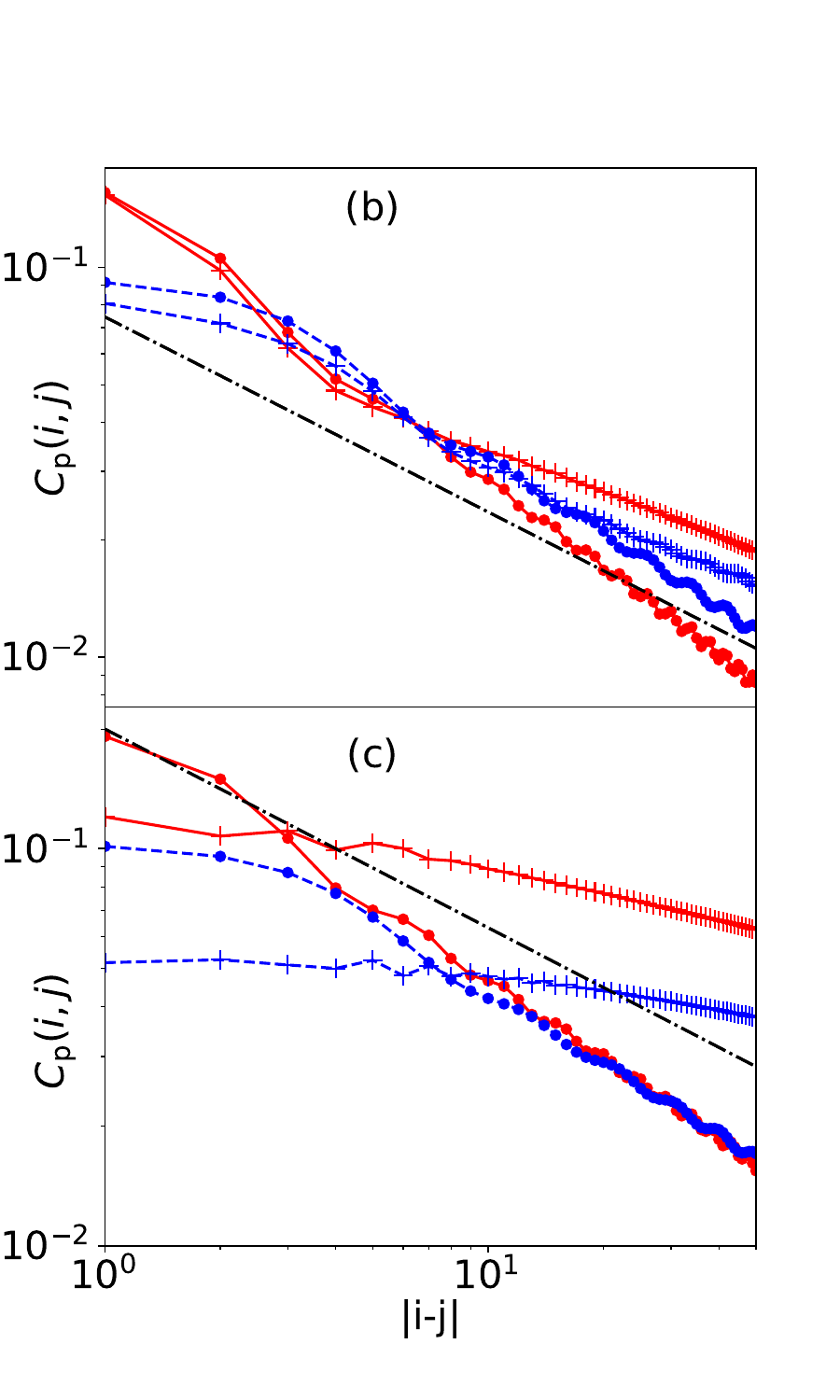}
\caption{ 
\textbf{(a)} Overview of bilayer system.
\textbf{(b)} and \textbf{(c)} show $C_{\rm p}$ (see text) when $k_F^{\rm p}$ and $k_F^{\rm m}$ are comparable (red crosses) and far apart (blue crosses).
Circles show $C_{\rm p}$ for isolated P-layer at matching densities.
Dash-dotted lines illustrate the slowest decay rate possible for the isolated {\rm P}-layer, at $K_{\rm p}=2$.
\textbf{(b)} ${U/t_{\rm p}=4}$, ${t_\perp/t_{\rm p}=0.4}$.
Red crosses show ${k^{\rm p}_F=0.81}$, ${k^{\rm m}_F=1.02}$, blue crosses show ${k^{\rm p}_F=0.39}$, ${k^{\rm m}_F=1.44}$ 
\textbf{(c)} ${U/t_{\rm p}=10}$, ${t_{\rm m}/t_{\rm p}=10}$, ${t_\perp/t_{\rm p}=3}$.
Red crosses show ${k^{\rm p}_F=0.68}$, ${k^{\rm m}_F=0.89}$, blue crosses show ${k^{\rm p}_F=0.32}$, ${k^{\rm m}_F=1.24}$
}
\label{pairing1}
\end{figure}
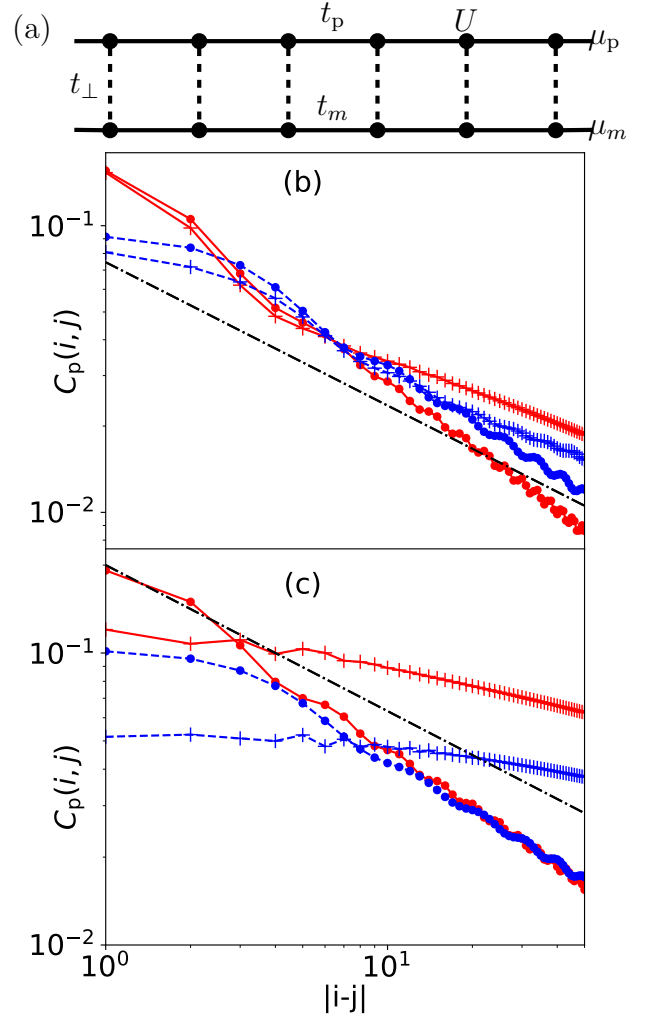
\begin{eqnarray}
\Delta_{\rm p} & = & \frac{1}{2}\left(\sum_{\sigma}E_{\rm GS,p}(N+2\sigma,\sigma)-2E_{\rm GS,p}(N,0)\right)
\end{eqnarray}
where $E_{\rm GS,p}(N,S_z)$ denotes the ground state energy of the isolated {\rm P}-layer with total charge $N$ and $z$-spin $S_z$.
We focus on two regimes, going up to $L=200$; unless stated otherwise, all data shown is for $L=200$~\cite{suppmat}:
\begin{eqnarray}
    \mbox{regime 1} & : & U/t_{\rm p} = 4, t_{\rm m}/t_{\rm p} = 1 \\
    \mbox{regime 2} & : & U/t_{\rm p} = 10, t_{\rm m}/t_{\rm p} = 10 
\end{eqnarray}
Unless noted otherwise, we work at $t_\perp/t_{\rm p}=0.4$ in regime 1, and at $t_\perp/t_{\rm p}=3.0$ in regime 2.
For regime 1, we find ${\Delta_{\rm p}/t_{\rm p}\approx0.69}$,
and for regime 2 it is ${\Delta_{\rm p}/t_{\rm p}\approx6.6}$ across the studied range of densities.
At temperature ${T=0}$, we use the SyTen-package based on matrix product states (MPS)~\cite{Schollwock2011,suppmat}, and at $T>0$ the ALF-package, based on auxiliary-field Quantum Monte Carlo (AFQMC)~\cite{Assaad2022,suppmat}.
At $T=0$, we thus drop $\mu_{\rm p}$ and fix the global density of electrons $n$, to $n/2=n_\uparrow=n_\downarrow=0.58$ for regime 1, and to $n/2=n_\uparrow=n_\downarrow=0.5$ for regime 2, adjusting the relative densities in each layer via $\mu_{\rm m}$.
We assign a nominal Fermi wave vector ${k_F^\lambda=\pi n_\lambda/2}$ to each layer, and by tuning $\mu_{\rm m}$ we explore the impact of having nesting, ${k_F^{\rm p}\approx k_F^{\rm m}}$, compared against violating nesting to various degrees, ${k_F^{\rm p}\neq k_F^{\rm m}}$.
At $T>0$, where densities are only fixed on average, we then tune both $\mu_{\rm p}$ and $\mu_{\rm m}$ so as to work at specific densities.
These are ${n_{\rm p}=0.513}$ and ${n_{\rm m}=0.647}$ for regime 1, and ${n_{\rm p}=0.245}$ and ${n_{\rm m}=0.753}$ for regime 2

Regime 1 is the 1D analogue to the basic setups studied previously in 2D, when tuning the properties of the metal to enhance superconductivity. 
Regime 2 has not previously been studied, and for 2D systems  meaningful lattice sizes would not be possible anyway, due to the much smaller linear dimensions accessible to 2D numerics~\cite{Zujev2014,Dee2022,Zhang2025}.
The purpose of regime~2 then is to maximally stabilise superconductivity in the {\rm P}-layer by using large values for both $t_{\rm m}$ and $U$.
The large $U$-value compensates for the weakening of effective pairing in the ${\rm P}$-layer due to the large $t_{\rm m}$-value (a reverse proximity effect).
Viewed the other way around, the high $t_{\rm m}$-value aims at reducing the single-particle gap induced in the metal by the {\rm P}-layer (proximity effect)~\cite{suppmat}. 
This unavoidable gap fundamentally changes the pair-pair coupling that the metal can mediate within the {\rm P}-layer:
in previous perturbative analytical treatments, a single-particle correlator ${\langle \hat{c}^\dagger_{i,\sigma,{\rm m}}(\tau) \hat{c}_{j,\sigma,{\rm m}}(0)\rangle}$ with algebraic decay in either space or imaginary time would mediate long-range pair-pair-coupling within the {\rm P}-layer which allows the system to side-step the Mermin-Wagner theorem (MWT) and can thus lead to a 1D state with full SC order~\cite{Lobos2009}.
Our non-perturbative numerics show that this analysis neglects the crucial impact of the proximity effect on the {\rm M}-layer~\cite{suppmat}.
The resulting gap for single-particle excitations causes the metal to mediate exponentially decaying pair-pair coupling in the {\rm P}-layer instead, which thus does not escape the MWT.
Still, this coupling can lead to a SC susceptibility enhanced to be far above the isolated {\rm P}-layer.
In particular, increasing $t_{\rm m}$ as done in regime 2 yields a longer-ranged pair-pair-coupling  within the confines of  exponential decay, resulting in strong improvements to SC susceptibilities. 

For both layers, we analyse the $s$-wave pair-pair and single-particle correlation functions respectively:
\begin{eqnarray}
    C_{\lambda}(i,j) & = &  \left\langle  \hat{c}^{\dagger}_{i,\uparrow,\lambda} \hat{c}^\dagger_{i,\downarrow,\lambda} \hat{c}_{j,\downarrow,\lambda} \hat{c}_{j,\uparrow,\lambda} \right\rangle ,\\
    S_{\lambda}(i,j) & = &  \left\langle  \hat{c}^{\dagger}_{i,\uparrow,\lambda}  \hat{c}_{j,\uparrow,\lambda} \right\rangle
\end{eqnarray}
We find that the maxima of $C_{\lambda}(i,j)$ can be fitted with ${\left[ \pi A_{C,\lambda}(\beta)/\left( \xi_{C,\lambda}(\beta) \sinh(\pi |i-j|/\xi_{C,\lambda}(\beta))\right)\right]^{K_{\lambda}^{-1}}}$, with ${\beta=T^{-1}}$, similar to the isolated Hubbard chain~\cite{Thierrybook2003}, but the amplitudes $A_{C,\lambda}$ are now temperature-dependent (see below).
The parameter $K_{\lambda}$ is a Tomonaga-Luttinger-liquid (TLL) parameter that encodes the interactions and controls the SC susceptibility, which algebraically diverges with the power ${K_{\lambda}^{-1}-2}$, as well as the finite size condensate peak at zero momentum, which scales as ${L^{1-K_{\lambda}^{-1}}}$~\cite{Thierrybook2003}.
As discussed above, for the maxima of $S_\lambda$ we have to use the ansatz ${A_{S,\lambda}e^{-|i-j|/\xi_{S,\lambda}}}$ for \textit{both} layers. 
The length scale $\xi_{S,{\rm m}}$ encodes the effective range of the pair-pair coupling induced in the {\rm P}-layer by the metal - the larger, the better the metal can stabilise the phase of pairs across the P-layer~\cite{suppmat}.
Conversely, the length scale $\xi_{S,{\rm p}}$ encodes the effective strength of pairing in the ${\rm P}$-layer, which at ${t_\perp>0}$ will depend not just on $t_{\rm p}$, $U$ and $\mu_{\rm p}$, but on all parameters of the system - the larger $\xi_{S,{\rm p}}$, the weaker the effective pairing active within in the P-layer.
All fits are performed for distances up to ${|i-j|=L/4}$ in order to avoid boundary effects that become pronounced beyond this point, with $i$ being one of the two cental sites in each layer.
\begin{figure}[htp!]
\centering
\hspace*{-0.4cm}\includegraphics[clip,trim= 0cm 0cm 0cm 0cm,scale=0.48]{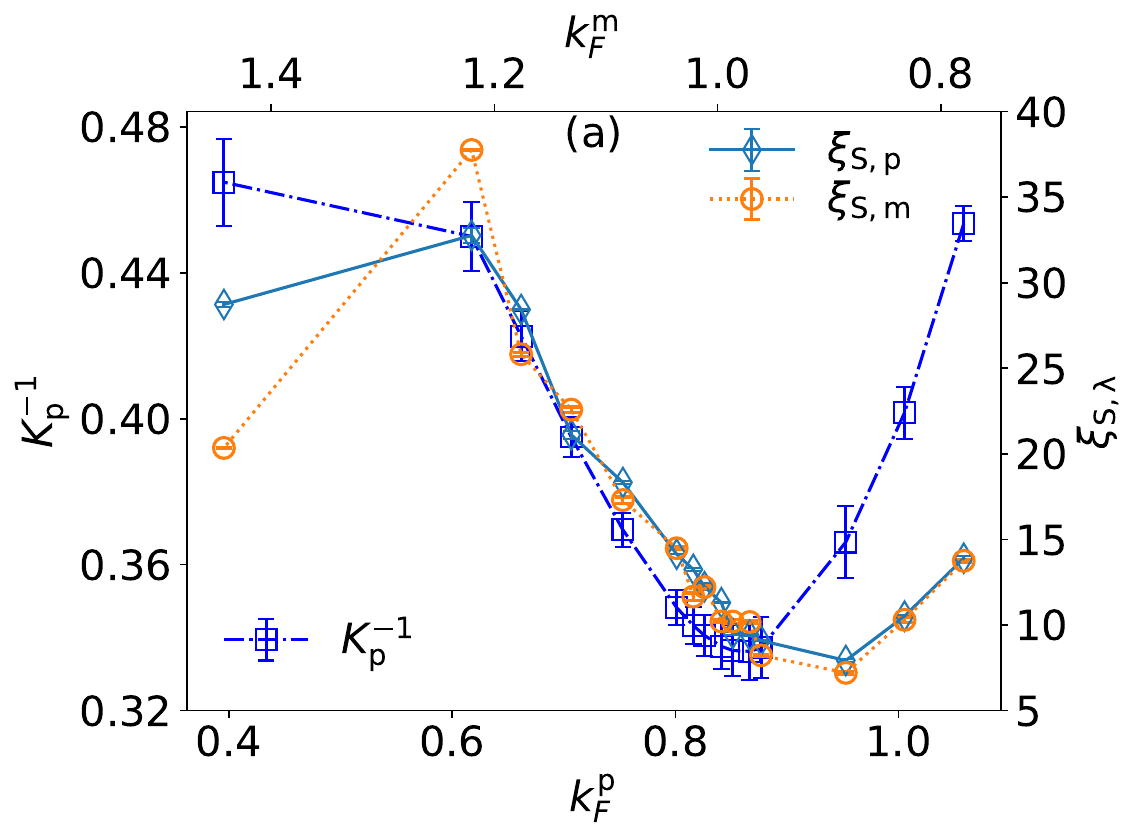}
    \includegraphics[scale=0.43]{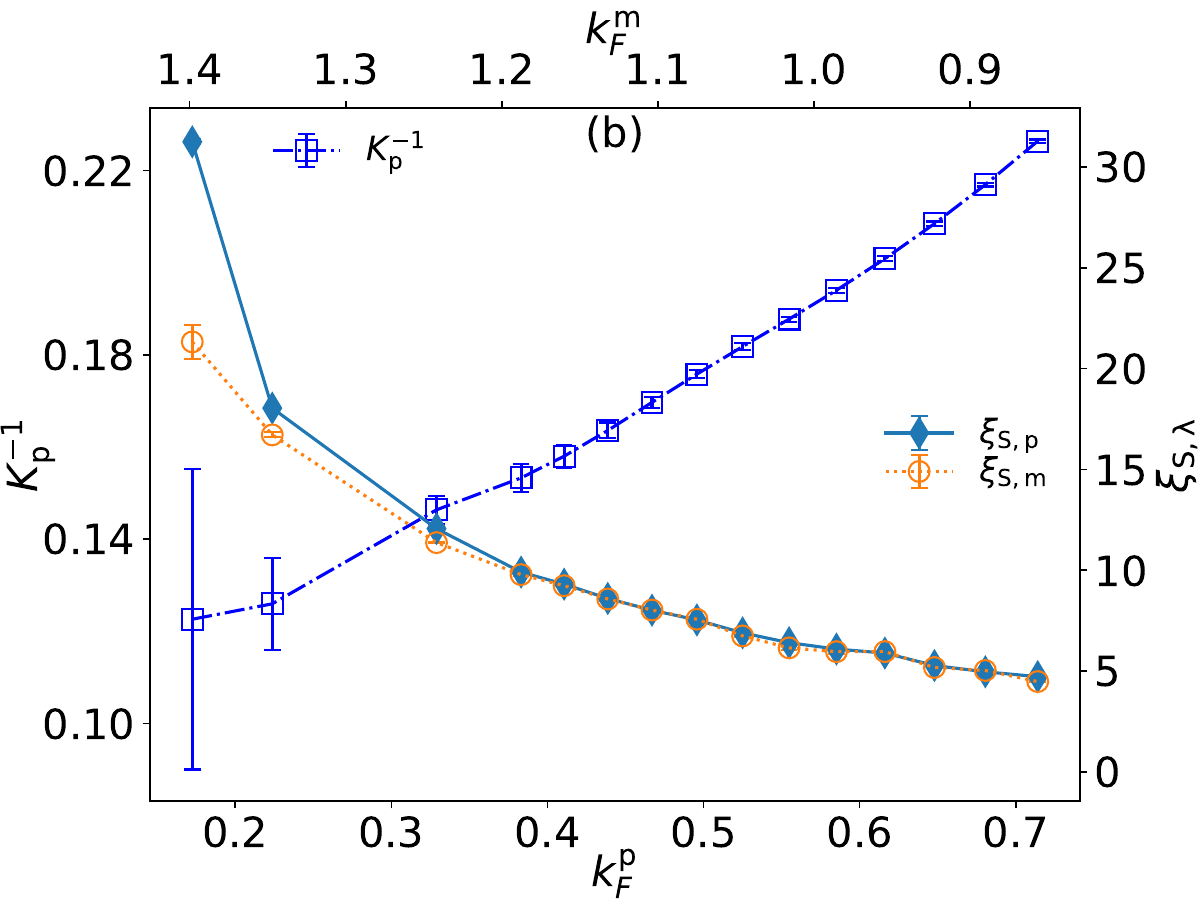}
    \includegraphics[scale=0.43]{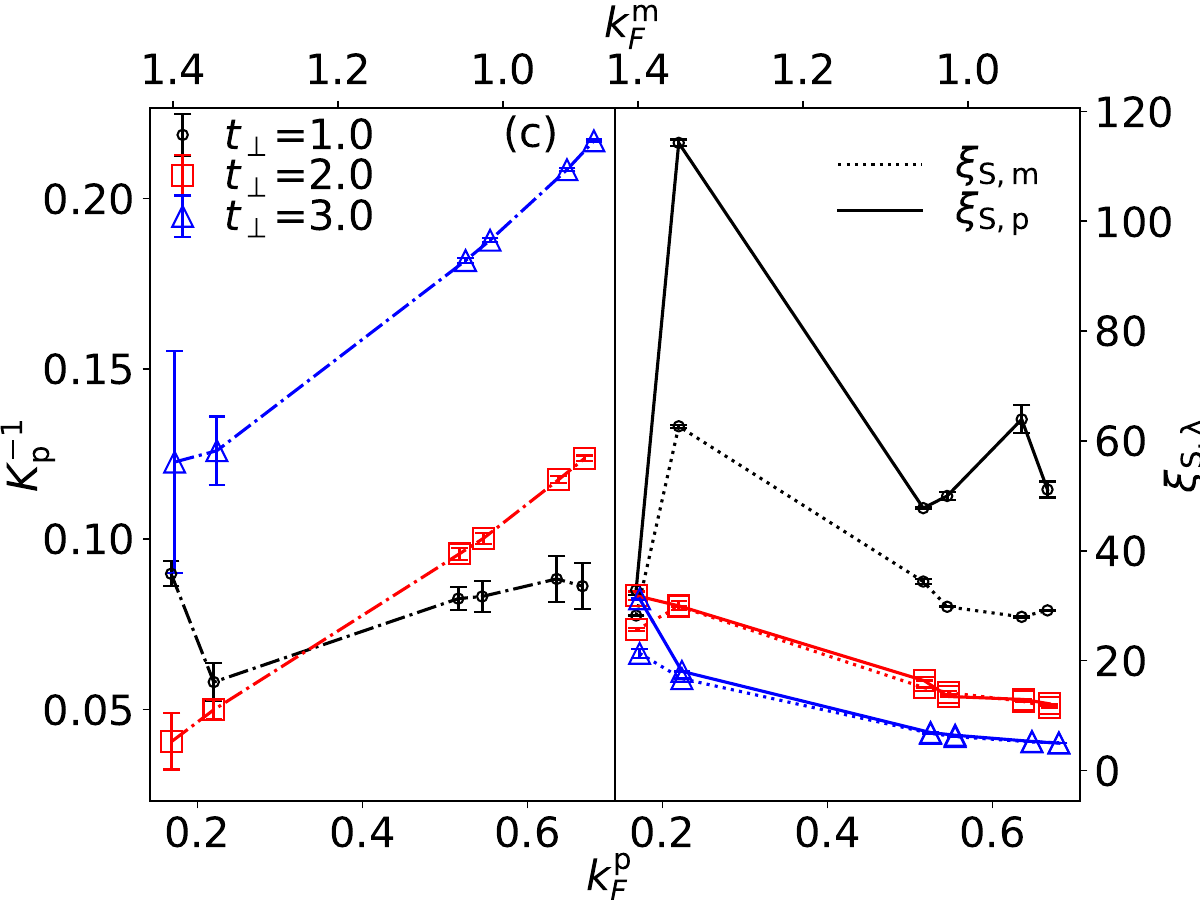}
   \caption{
   Left $y$-axis: $K_{\rm p}^{-1}$ vs. $k_F^{\rm p}$.
   Right $y$-axis: $\xi_{S,\rm p}$ and $\xi_{S,\rm m}$ vs. $k_F^{\rm p}$.
   The top axis allows to read off the corresponding $k_F^{\rm m}$-values.
   \textbf{(a)} Regime 1
   \textbf{(b)} Regime 2
   \textbf{(c)} Regime 2 for $t_\perp/t_{\rm p}=1$ (black), 
   $t_\perp/t_{\rm p}=2$ (red), 
   $t_\perp/t_{\rm p}=3$ (blue). 
   }
   \label{U10_coefficients}
\end{figure}

As illustrated in~\cref{pairing1}b, c, both regimes can outperform the isolated Hubbard chain significantly, which is limited to $K_{\rm p}^{-1}>1/2$~\cite{Thierrybook2003}. 
In fact, for the 1D hybrid systems we find $K_{\rm p}^{-1}<1/2$ for all studied parameters (c.f.~\cref{U10_coefficients}).
Especially regime 2 realizes remarkable enhancements to pair-par-correlations and thus $K_{\rm p}$.
Not only is $K_{\rm p}$ well above $2$ in this regime, but it can be up to a factor of $4$ larger than the $K_{\rm p}$ of the isolated {\rm P}-layer at comparable densities. 
We also note that we even find values of $K_{\rm p}$ much higher than the theoretical optimum of $K_{\rho,+}=4$ for the symmetric charge mode of two weakly coupled identical chains which both have pairing.
Especially the high $K_{\rm p}$-values achievable in regime 2 will cause any observable, such as condensate peaks, to scale as barely distinguishable from an ordered SC state at the $L$-values amenable to numerics.

Characterizing the ground states of our model across a wide range of relative densities $n_{\rm p, m}$ and thus Fermi wave vectors $k_F^{\rm p, m}$, a striking result is that we find ${\xi_{S,{\rm p}}\approx\xi_{S,{\rm m}}}$ almost across the board, as shown in~\cref{U10_coefficients}.
Thus, even in the regime where interlayer tunneling $t_\perp$ is the weakest coupling scale and is well below $\Delta_{\rm p}$, both layers with their very different internal Hamiltonians still attempt aligning the correlation lengths associated with gapped single-electron excitations.
However, in regime 2, when ${k_F^{\rm p}\ll k_F^{\rm m}}$ at ${t_\perp/t_{\rm p}=2.0, 3.0}$, and throughout for ${t_\perp/t_{\rm p}=1.0}$, the system does not succeed at aligning $\xi_{S,{\rm p}}$ and $\xi_{S,{\rm m}}$ (c.f. ~\cref{U10_coefficients}c).

Even more striking is a mechanism for tuning system performance that is uncovered by the data in~\cref{U10_coefficients}:
controlling the positions of the wave vectors $k_F^{\rm p, m}$ relative to each other allows to optimise the SC susceptibilities of both regimes, i.e. to find the best possible balance between maximising  the range of pair-pair coupling, i.e. increasing $\xi_{S,{\rm m}}$, and maximising pairing in the {\rm P}-layer, i.e. minimising $\xi_{S,{\rm p}}$.
For regime 1 (c.f.~\cref{U10_coefficients}a), where the mobility of pairs is already high in the isolated {\rm P}-layer, that optimal point sits close to, though not quite at, the nesting condition ${k_F^{\rm p}= k_F^{\rm m}}$.
This minimizes the loss of pairing energy due to the reverse proximity effect, sitting near the minimum of the $\xi_{S,{\rm p}}(k_F^{\rm p})$-curve.
Conversely, due to the high intrinsic mobility of pairs within the {\rm P}-layer, a low level of additional support stemming from the metal-mediated pair-pair coupling is sufficient to achieve the optimal SC susceptibility, with $\xi_{S,{\rm m}}$ sitting only a little above the minimum of the $\xi_{S,{\rm m}}(k_F^{\rm p})$-curve.
For regime 2, the system improves its SC susceptibility in the opposite manner, by detuning $k_F^{\rm p}$ and $k_F^{\rm m}$ as much as possible, which increases $\xi_{S,{\rm p}}$ and $\xi_{S,{\rm m}}$, thus respectively depressing the strength of pairing and increasing the range of pair-pair-coupling that the metal mediates in the {\rm P}-layer.
With a low intrinsic pair-mobility of the {\rm P}-layer due to the large $U$-value, sacrificing pairing energy is a small price to pay in return for the strong enhancement to pair-mobility afforded by an increasing $\xi_{S,{\rm m}}$ as $k_F^{\rm p}$ and $k_F^{\rm m}$ are tuned away from each other.
In this way, tuning $k_F^{\rm p,m}$ and $t_{\rm m}$ allows each regime to get the most out of the metallic reservoir according to their specific needs.
Lowering the value of $t_\perp$ from the default ${t_\perp/t_{\rm p}=3}$ chosen for regime 2 indicates that tuning $k^{\rm p}_F$ could allow access to two different regimes, as shown in~\cref{U10_coefficients}c:
at low $k^{\rm p}_F$, the SC susceptibility peaks in-between ${t_\perp/t_{\rm p}=1}$ and ${t_\perp/t_{\rm p}=3}$, while at higher $k^{\rm p}_F$ susceptibility decreases monotonically across the three ${t_\perp/t_{\rm p}}$-values studied.
Moreover, at constant ${t_\perp/t_{\rm p}}$ the results summarized in~\cref{U10_coefficients}c confirm our finding that the growing range of pair-par coupling mediated by the metal, and weakening of pairing in the P-layer, both caused by detuning $k^{\rm p}_F$ and $k^{\rm m}_F$ relative to each other, largely serves to improve SC susceptibility.
Yet, we also find examples where this is not the optimal way, namely at low $k^{\rm p}_F$ and at intermediate ${t_\perp/t_{\rm p}=2}$.
Here, the system can only further lower $K_{\rm p}^{-1}$ by rebalancing in favour of a slightly strenghtened pairing and reduced coupling range.
The data in~\cref{U10_coefficients}c further illustrates that the range of pair-pair coupling, as parametrized by $\xi_{S,{\rm m}}$, is not the only factor determining the efficacy of metal-induced stabilization of superconductivity.
This is evidenced by the systems with higher $t_\perp$ often outperforming those with lower $t_\perp$ regarding SC susceptibility, even though these show much larger $\xi_{S,{\rm m}}$.
This can be understood by looking at the leading order of the metal-mediated pair-pair coupling, which is controlled by a prefactor of 
${t_\perp^4/(\tilde{\Delta}_{\rm p}+\Delta_{\rm m})^2}$ (c.f.~\cite{Lobos2009}), where $\tilde{\Delta}_{\rm p}$ denotes the renormalized pairing gap of the P-layer, and $\Delta_{\rm m}$ is the gap induced in the metallic layer by the proximity effect.

\begin{figure}[hbt!]
\includegraphics[scale=0.42]{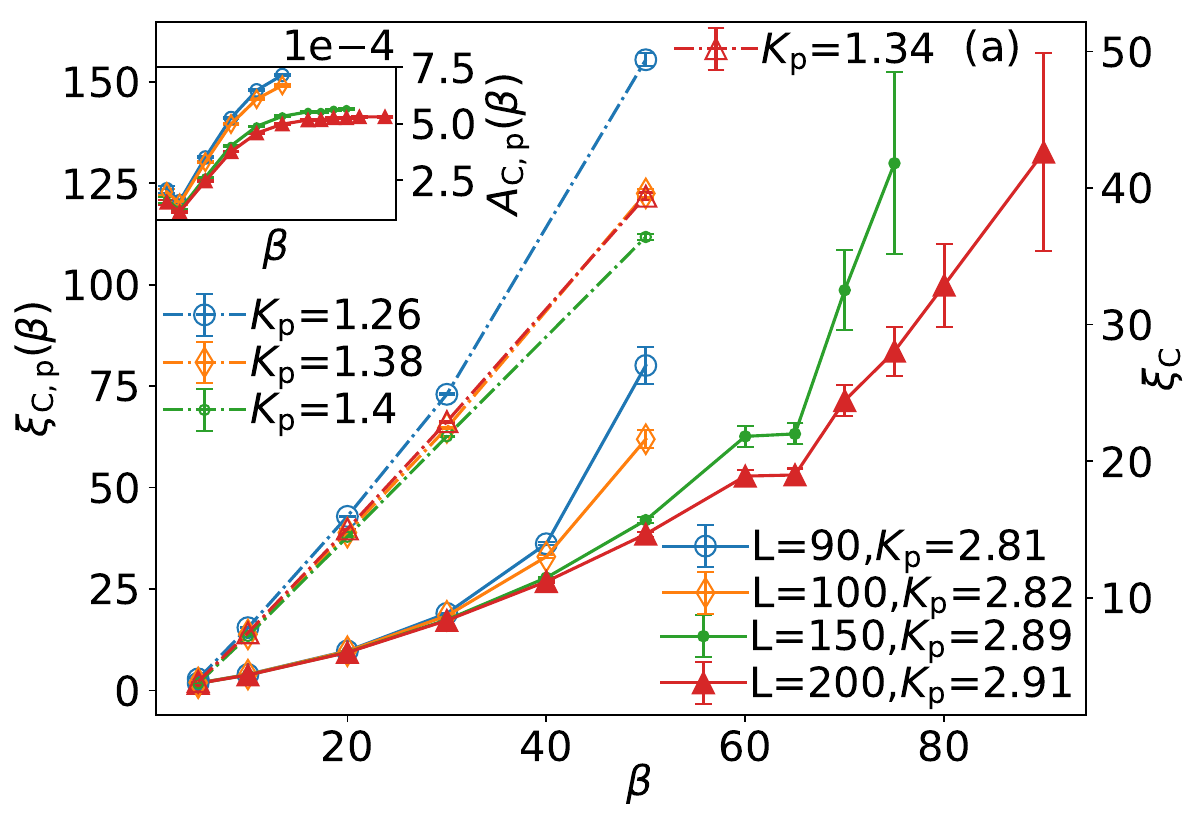}
\includegraphics[scale=0.42]{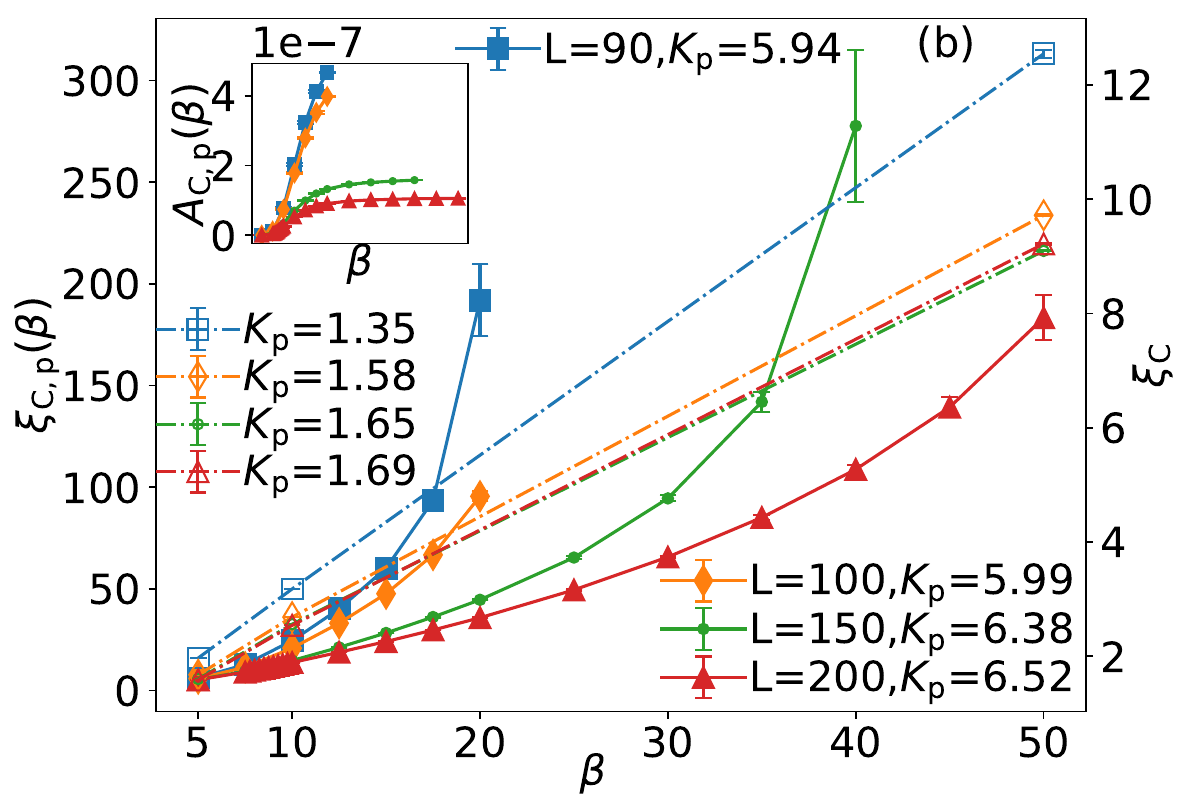}
    \caption{
    Thermal SC correlation length $\xi_{\rm C ,\lambda}$ (left $y$-axis) versus $\beta$ at various $L$-values, extracted from fitting ALF-data.
    Insets show the corresponding fitted values of the non-universal amplitudes $A_{C,{\rm p}}$.
    The values of $K_{\rm p}$ shown have been extracted from independent MPS-based calculations at $T=0$.
    We see strong superlinear scaling of $\xi_{\rm C ,{\rm p}}$ in $\beta$ relative to the isolated baseline $\xi_{C}$ (right $y$-axis).
    \textbf{(a)} Regime 1 at ${k_F ^{\rm p}=0.806}$ and ${k_F^{\rm m}= 1.017}$ 
    \textbf{(b)} Regime 2 at ${k_F^{\rm p}=0.3848}$,  ${k_F^{\rm m}=1.1844}$ 
    }
    \label{length_scale_U10}
\end{figure}
\FloatBarrier

The strong boosts to the superconducting properties of the P-layer by the metal show up at ${T>0}$ as well, which is important for designing improved hybrid SC devices in practice.
As shown in~\cref{length_scale_U10}, we find that the thermal SC correlation length ${\xi_{C,{\rm p}}(\beta)}$ is not only far larger than for the isolated P-layer, it now scales superlinearly in $\beta$, where the isolated P-layer exhibits linear scaling in accordance with the predictions of conformal field theory.
The only consistent way of obtaining ${\xi_{C,{\rm p}}(\beta)}$ from fits is to assume the non-universal amplitudes ${A_{C,{\rm p}}}$ to be temperature-dependent, which they are not otherwise~\cite{Thierrybook2003}.
This expresses the fact that the form of any microscopic action derived for the P-layer will depend on the state of the metal, which in turn will change with $\beta$.
With ${\xi_{C,{\rm p}}(\beta)}$ boosted by more than an order of magnitude over that of the isolated P-layer, with growing $\beta$ the thermal SC correlation length rapidly matches and exceeds the central half of the system, i.e. that portion which exhibits bulk behaviour undistorted by boundary effects and thus the region on which the fitting is performed.
This manifests as an apparent onset of a divergence in ${\xi_{C,{\rm p}}(\beta)}$ once this lengthscale grows beyond roughly $L/4$, as the finite-size system becomes practically indistinguishable from its state at $T=0$ from there on.
In contrast to the isolated P-layer, the size of bilayer systems can thus be grown superlinearly with $\beta$ and still appear to be entirely within the low- or zero-temperature regime.

This work demonstrates the powerful benefits to SC properties that can be derived from a metallic reservoir, and especially from  tuning the parameters of the metal such as to deliberately promote mediated pair-pair coupling in conjunction with the relative positions of the two layer's nominal Fermi surfaces.
It thus provides a framework on how to approach higher-dimensional analogs such as 2D bilayer systems in which these effects - and their impact on the BKT transition temperature - have not yet been explored on purpose, either experimentally or theoretically, but that would be especially important in the quest for a targeted design of high-$T_c$ SC devices.
Towards the latter, large-scale AFQMC-based simulations could be one possible approach, as would be CT-INT-based QMC algorithms that integrate out the metal~\cite{Gull2011}.
Furthermore, in the regime where tunneling in the $y$-direction is weak in both layers, the recently developed MPS+MF technique for fermions~\cite{Bollmark2023} would make it possible to study far larger 2D bilayer systems than those amenable to any practical QMC-based simulations, and thus to probe regimes with large values of $t_{\rm m}/t_{\rm p}$ and / or low $U/t_{\rm p}$ which might show exceptional enhancement to SC properties.

This work was supported by an ERC Starting Grant from the European Union’s Horizon 2020 research and innovation programme under grant agreement No. 758935; and the UK’s Engineering and Physical Sciences Research Council [EPSRC; grant number EP/W022982/1]. 
This work is also supported by the Swiss National Science Foundation under grant number 200020-219400.
The computations were enabled by resources provided through multiple EPSRC ``Access to HPC'' calls (Spring 2023, Autumn 2023, Spring 2024 and Autumn 2024) on the ARCHER2, Peta4-Skylake and Cirrus compute clusters, as well as by computer time awarded by the National Academic Infrastructure for Supercomputing in Sweden (NAISS).
This work was supported by a grant from the Swiss National Supercomputing Centre (CSCS) under project ID s1307 on Alps.
The authors also acknowledge the use of the HWU high-performance computing facility (DMOG) and associated support services in the completion of this work.

\bibliography{references}

\end{document}


\title{Supplementary material \\ Strong enhancements to superconducting properties of 1D systems from metallic reservoirs}

\author{J. E. Ebot}
\affiliation{SUPA, Institute of Photonics and Quantum Sciences, Heriot-Watt University,
Edinburgh EH14 4AS, United Kingdom}

\author{Sam Mardazad}
\affiliation{SUPA, Institute of Photonics and Quantum Sciences, Heriot-Watt University,
Edinburgh EH14 4AS, United Kingdom}

\author{Lorenzo Pizzino}
\affiliation{DQMP, University of Geneva, 24 Quai Ernest-Ansermet, 1211 Geneva, Switzerland}

\author{Johannes S. Hofmann}
\affiliation{Department of Condensed Matter Physics, Weizmann Institute of Science, Rehovot 76100, Israel}
\affiliation{Max-Planck-Institut f\"ur Physik komplexer Systeme,N\"othnitzer Strasse 38, 01187 Dresden, Germany}

\author{Thierry Giamarchi}
\affiliation{DQMP, University of Geneva, 24 Quai Ernest-Ansermet, 1211 Geneva, Switzerland}

\author{Adrian Kantian}
\affiliation{SUPA, Institute of Photonics and Quantum Sciences, Heriot-Watt University,
Edinburgh EH14 4AS, United Kingdom}

\date{\today}

\maketitle

\appendix
\section{Performance of DMRG for ground state calculations}
Here, we provide details of our DMRG calculations, along with results for other systems sizes below $L=200$.
DMRG is based on a one-dimensional ansatz wavefunction that is variationally optimized via repeated iterations.
We are using the version of this algorithm based on matrix product state (MPS), as implemented in the SyTen codebase, whose accuracy  is controlled by the so-called bond dimension $m$~\cite{Schollwock2011}.
For the present work, we have calculated the ground-states of our systems with open boundary conditions (OBC) and kept up to ${m=8024}$ states. 
At this bond dimension, we have a truncated weight (sum of the discarded components of the wave function at each iterative steps) of $\mathcal{O}(10^{-14})$ to $\mathcal{O}(10^{-16})$ at most. 
We have used two-site updates throughout, and used $8$ Krylov-states per update.
We stopped sweeping when the change in energy after one sweep dropped below $10^{-16}$.

\section{Proximity-effect in the metal}
\begin{figure}
\centering
\includegraphics[width=1.1\linewidth,trim= 0 0 -10 20]{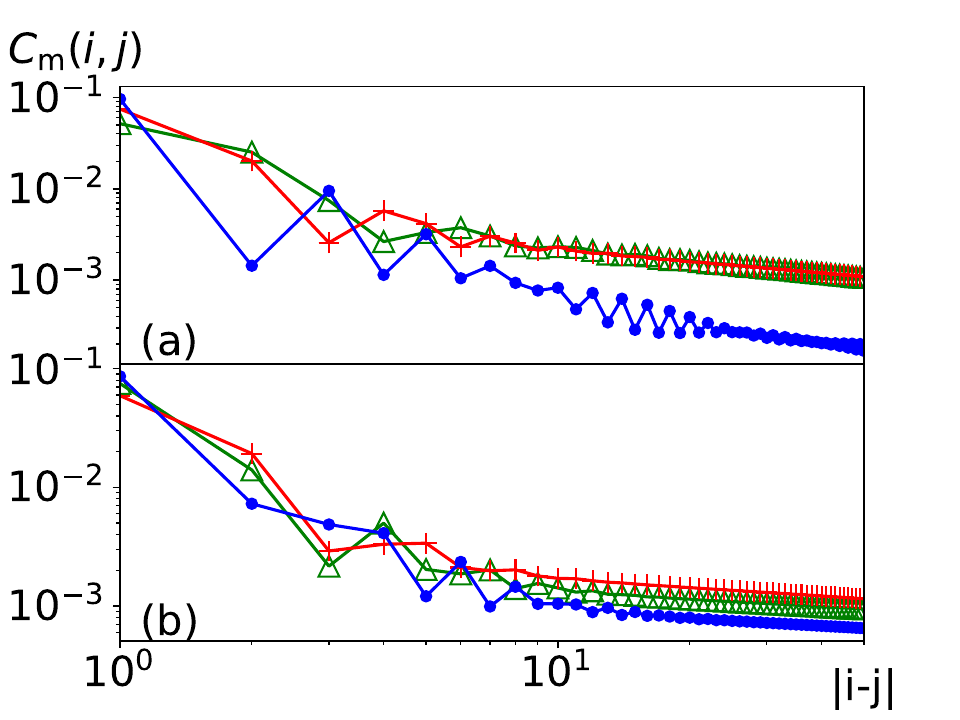}
\caption{Metallic pair correlation functions for different densities with $L=200$. 
(a) regime 1; ${k_F^{\rm p}=1.058}$, ${k_F^{\rm m}=0.779}$ (\protect\opentriangle),${k_F^{\rm p}=0.816}$, $k_F^{\rm m}=1.022$ (\protect\elineplus), and ${k_F^{\rm p}=0.395}$, ${k_F^{\rm m}=1.442}$ (\protect\elinecircle). 
(b) regime 2;
${k_F^{\rm p}=0.496}$, ${k_F^{\rm m}=1.075}$ (\protect\opentriangle), ${k_F^{\rm p}=0.681}$, ${k_F^{\rm m}=0.89}$ (\protect\elineplus) and ${k_F^{\rm p}=0.329}$, ${k_F^{\rm m}=1.242}$ (\protect\elinecircle).
}
\label{metallic_pairing}
\end{figure}

At any finite value of $t_\perp$, the P-layer will induce pairing in the metallic layer. 
Such a proximity effect opens a gap to spin excitations in the M-layer and thus leads to pair-pair correlations $C_{\rm m}(i,j)$ decaying significantly slower than for free electrons.
In \cref{metallic_pairing}, we illustrate this behavior for several pairs of $k_F$-values drawn from the range of parameters shown in fig. $1$b and fig. $1$c in the main text.
At the same time, single-electron propagation is changed fundamentally compared to the isolated metal, from being long-ranged, i.e. algebraically decaying  (in both space and imaginary time), to short-ranged, i.e. exponentially decaying, the envelope of which is controlled by $\xi_{S,{\rm m}}$.
Thus, the M-layer mediates extended-range pair-pair coupling within the P-layer via both single-electron propagation - as proven in the main text via the dependence of $K_{\rm p}$ on $\xi_{S,{\rm m}}$ - as well as via pair-propagation within the M-layer.
The data shown in \cref{metallic_pairing} then confirms physical intuition about how these two stabilization mechanisms are in competition with each other:
the pair-pair correlations in the M-layer are weakened by detuning $k_F^{\rm p}$ and $k_F^{\rm m}$, which however improves $\xi_{s,{\rm m}}$ and thus stabilization via independently propagating electrons.

\section{Using $t_\perp$ to improve the trade-off between pairing amplitude and superconducting stiffness}
\begin{figure}[hbt!]
    \centering
     \includegraphics*[width=\linewidth]{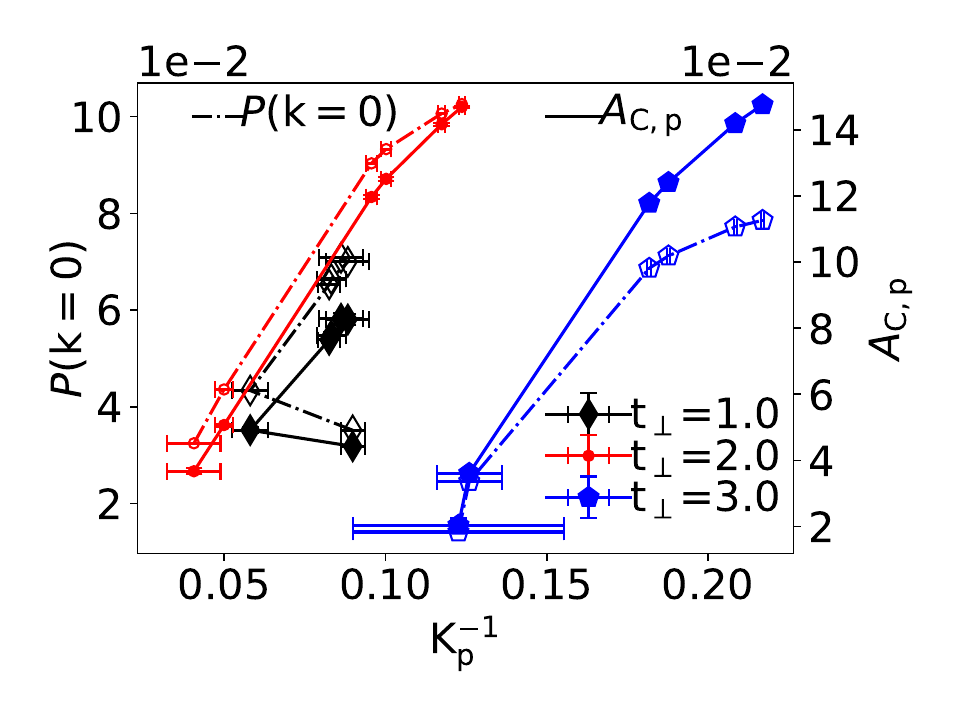}
    \caption{ 
    $P(0)$ vs. $K^{-1}_{\rm p}$ (full symbols) and $A_{C,{\rm p}}(\beta)$ at $T=0$ vs. $K^{-1}_{\rm p}$ (open symbols) in regime 2 at $t_\perp/t_{\rm p}=1.0, 2.0, 3.0$. The lines are to guide the eye.
    }
    \label{p_0}
\end{figure}

Here, we provide more results for different values of $t_\perp$ in regime 2, which serve to illustrate the trade-off between pairing amplitude and superconducting stiffness, and the degree to which inter-layer coupling can alleviate this. 
At fixed $t_\perp$, a higher superconducting susceptibility and stiffness means low $K_{\rm p}^{-1}$. 
To visualize this trade-off, \cref{p_0} plots the superconducting pairing peak  $P(0)$ against $K_{\rm p}^{-1}$ for $L=200$, where $K_{\rm p}^{-1}$ is changed by tuning $k_F^{\rm p,m}$ relative to each other (c.f. main text).
Here, we use $P(k) =\dfrac{1}{L} \sum_{j=1}^{L} C(i,j) e^{ik*(rj-(L/2) )}$ to
provide information on the long-range order of $C_{\rm p}(i,j)$ and the strength of overall pairing at zero temperature. 
It thus serves an analogous role as $A_{C,{\rm p}}(\beta)$ does at finite temperature.
Using these measures, the data shown in \cref{p_0} demonstrates that an intermediate value of $t_\perp$ can achieve strong pairing and high superconducting stiffness.
\clearpage

\section{Size-dependence of DMRG results}
Here, we show results for all system sizes that we studied, $L=90, 100, 150, 200$.
Compared to the results shown in Fig. 2a of the main text, \cref{u4_coeff}a summarizes the data for all sizes in regime 1, while \cref{u10_coeff}a shows the same for regime 2.
We also show an approximate estimate of the spin gap $\Delta _{\rm m}$ in the metal in \cref{u4_coeff}b and \cref{u10_coeff}b for regimes 1 and 2 respectively.
This estimate is based on the correlation length $\xi _{\rm S,m}$ and the assumption that the speed of sound of the M-layer is not changed too much from its value at $t_\perp=0$:
\begin{equation}
\Delta_{\rm S,m} =\frac{2t_m \sin({k^{\rm m}_{F}})}{\xi _{\rm S,m}}
\end{equation}
\begin{figure}[hbt!]
    \centering
    \includegraphics*[width=\linewidth]{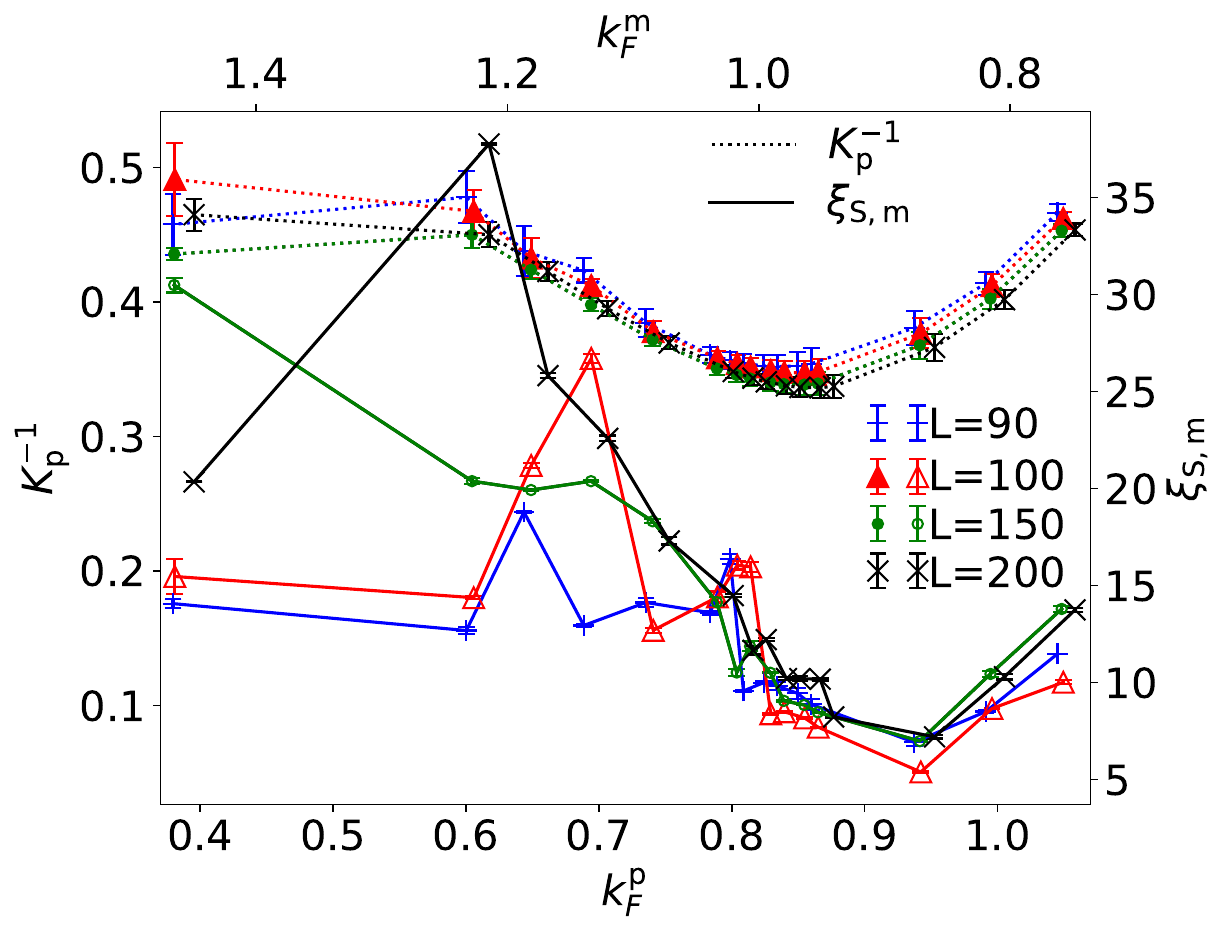}
\end{figure}
\begin{figure}[hbt!]
    \centering
 \hspace*{-1cm}\includegraphics*[width=0.95\linewidth]{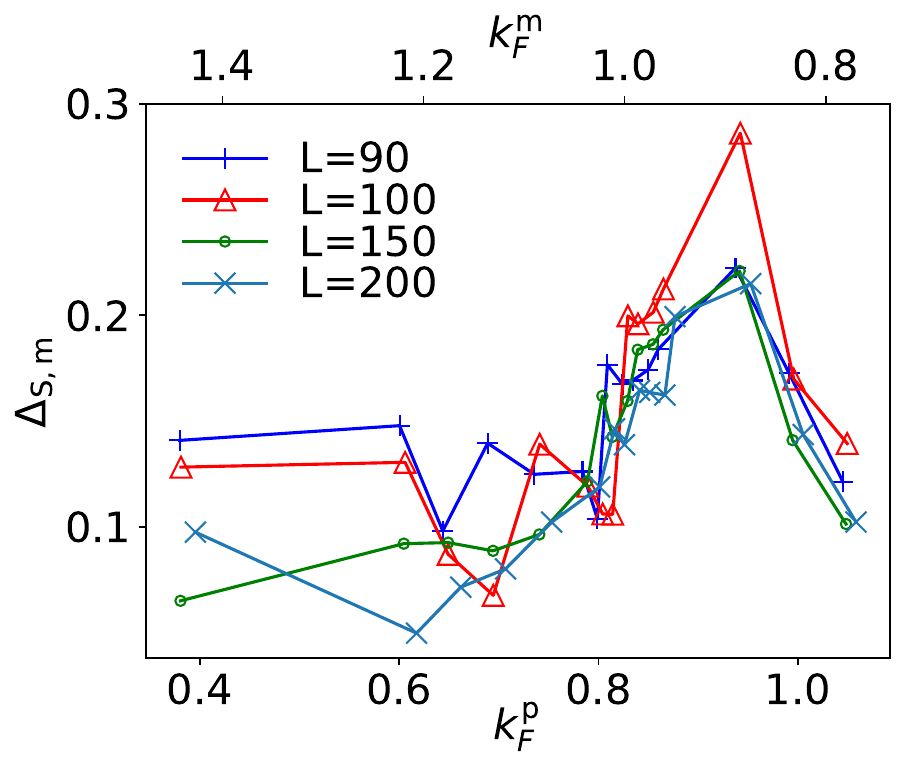}
    \caption{ 
    Regime 1: (top fig) $K^{-1}_{\rm p}$ and $\xi _{\rm S, m}$ vs. $k^{\rm p}_{F}$, (bottom fig) $\Delta _{\rm S, m}$ vs. $k^{\rm p}_{F}$, for different values of $L$. 
    The corresponding $k_F^{\rm m}$-values can be read off from the top x-axis. 
    }
    \label{u4_coeff}
\end{figure}
\begin{figure}[hbt!]
    \centering
     \includegraphics*[width=\linewidth]{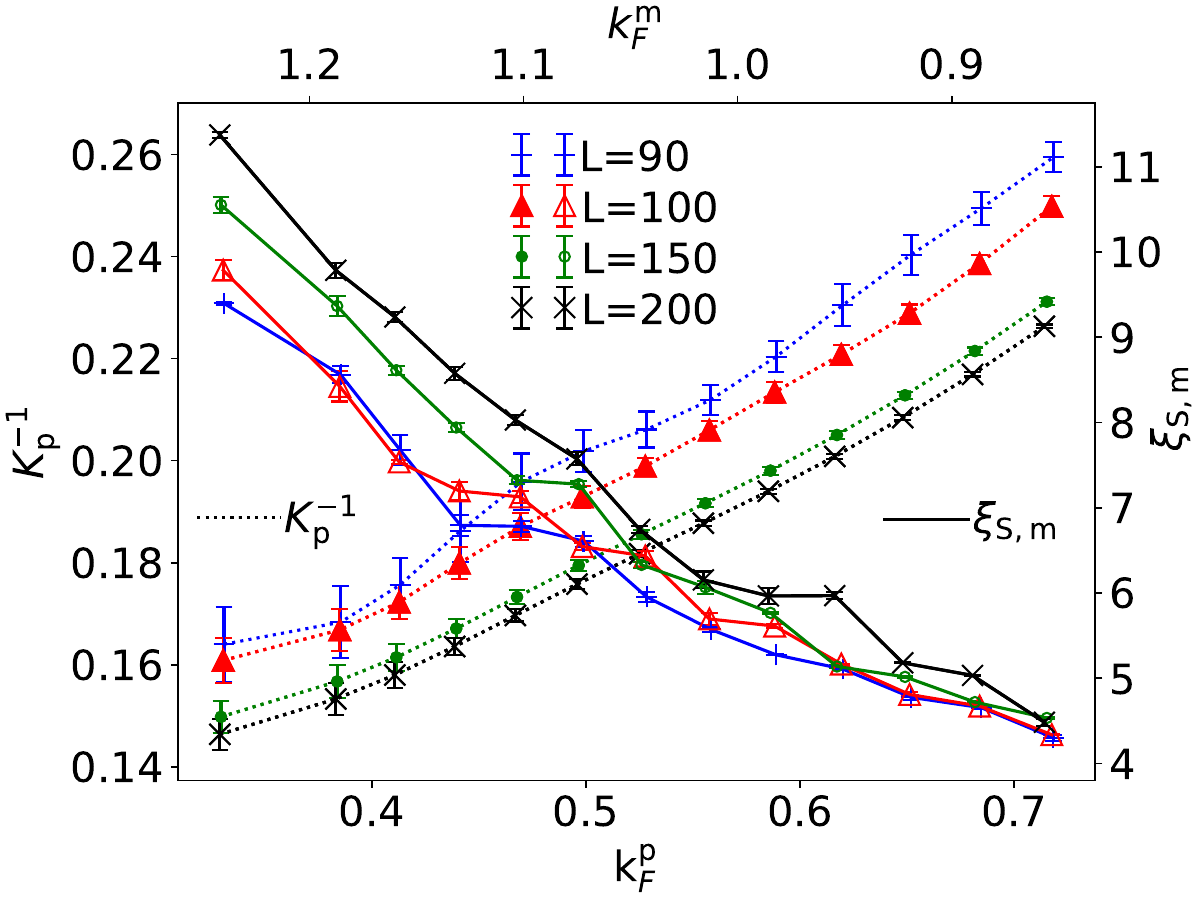}
\end{figure}
\begin{figure}[hbt!]
    \centering
         \hspace*{0.16cm}\includegraphics*[width=\linewidth]{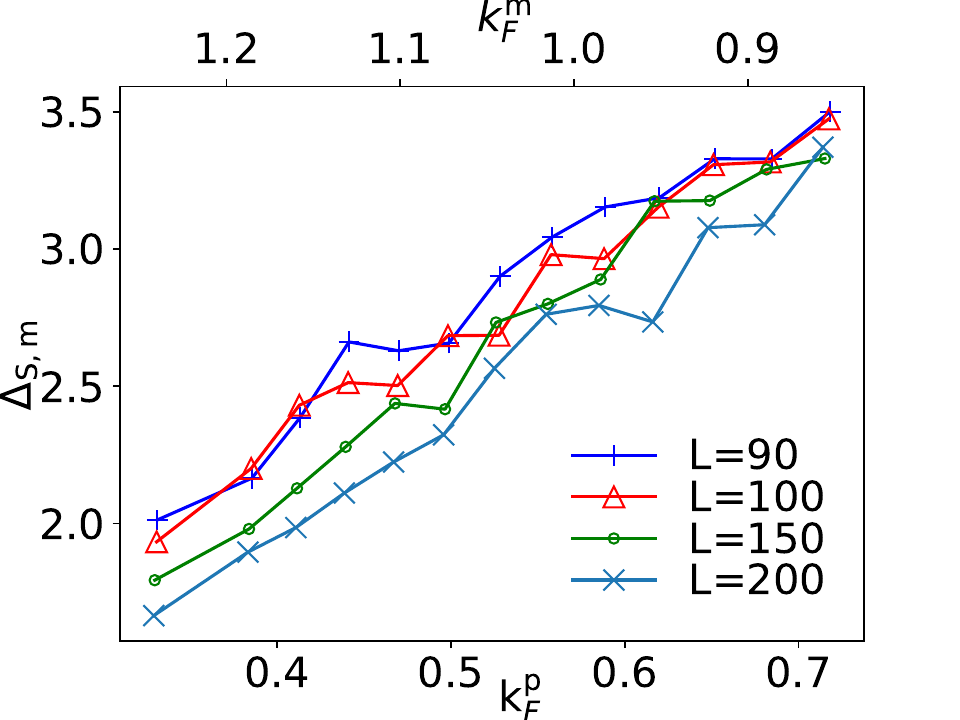}
    \caption{ Regime 2: (top fig) $K^{-1}_{\rm p}$ and $\xi _{\rm S, m}$ vs. $k^{\rm p}_{F}$, (bottom fig) $\Delta _{\rm S, m}$ vs. $k^{\rm p}_{F}$, for different values of $L$. 
    The corresponding $k_F^{\rm m}$-values can be read off from the top x-axis. }
    \label{u10_coeff}
\end{figure}
\section{Performance of AFQMC for finite-temperature calculations}
\begin{figure}[hbt!]
  \includegraphics[width=0.8\linewidth]{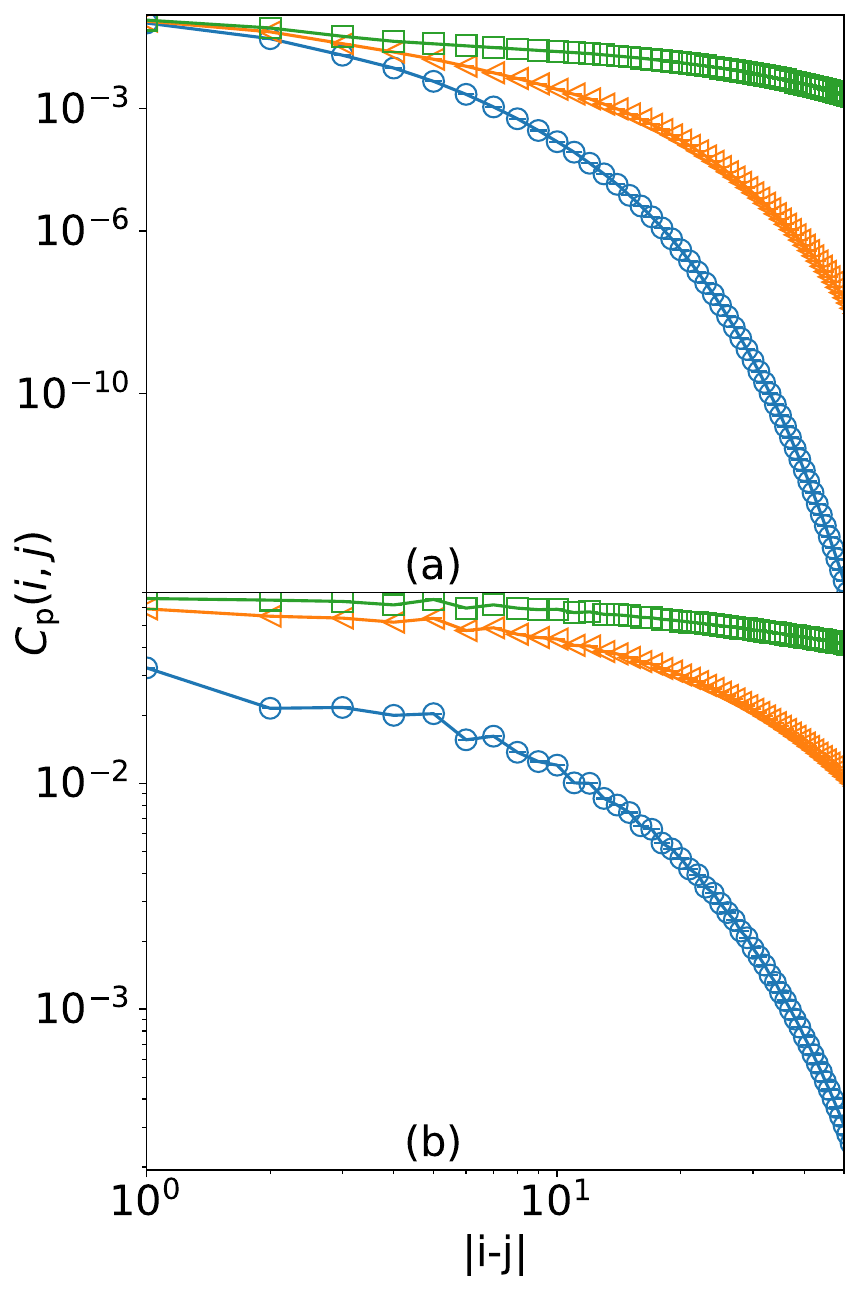}
   \caption{
   Pair-pair correlations function in the 
   P-layer, ${\beta=5}$ (\protect\opencircledash), ${\beta=10}$ (\protect\opentriangle), ${\beta=30}$ (\protect\opensquaredash). 
   (Top figure (a)) regime 1: with $n_{\rm p} \approx 0.512$, $n_{\rm m} \approx 0.647$  and (bottom figure (b)) regime 2: $n_{\rm p} \approx 0.245$, $n_{\rm m} \approx 0.754$. 
    }
    \label{finite-cor}
\end{figure}

We use the auxiliary field Quantum Monte Carlo (AFQMC) method as implemented in the ALF package ~\cite{Assaad2022} to compute the properties at finite-temperature thermal equilibrium. 
For the discretized imaginary time we have used a time step ${\Delta\tau=0.05}$ in regime 1 and ${\Delta\tau=0.02}$ in regime 2. 
In Fig.~\ref{finite-cor}, we show the pair-pair correlation functions $C_{\rm p}(i,j)$ in the pairing layer at different $\beta$ values, which illustrates the rapid onset of quasi-order across the system-bulk at surprisingly high temperatures.

Keeping the densities $n_{\rm p, m}$ in the finite-temperature AFQMC calculations consistent with changing temperatures requires tuning of the chemical potentials $\mu_{\rm p, m}$ of both layers, as summarized in table
\ref{table:1} and \ref{table:2} for both regimes.
\begin{table}[h!]
    \centering
    \begin{tabular}{|c|c|c|c|c|c|c|}\hline
                \multicolumn{7}{|c|}{Regime 1}  \\ \hline
                 $\beta$         & 5      & 10     &    20  & 30     &   40    &   50    \\  \hline
                 $\mu _{\rm p}$  & -0.594 & -0.604 & -0.613 & -0.6155 & -0.617 & -0.618  \\ \hline
                 $\mu _{\rm m}$  & -1.011 & -1.051 & -1.050 & -1.053  & -1.053 & -1.052   \\ \hline
    \end{tabular}
        \begin{tabular}{|c|c|c|c|c|c|c|} \hline
                 $\beta$         &   60     & 65     &  70     & 75       & 80      &  90 \\  \hline
                 $\mu _{\rm p}$   & -0.61813 & -0.6183 & -0.6185 & -0.6185 & -0.61855 & -0.61855 \\ \hline
                 $\mu _{\rm m}$   & -1.0509 & -1.051   & -1.051 & -1.051  &  -1.051  & -1.051  \\ \hline
    \end{tabular}
    \caption{}
    \label{table:1}
\end{table}
%
\begin{table}[h!]
    \centering
    \begin{tabular}{|c|c|c|c|c|c|c|c|}\hline
            \multicolumn{8}{|c|}{Regime 2}  \\ \hline
                 $\beta$         & 5      & 10     &    25    & 30       &   50    &  70    &  90 \\  \hline
                $\mu _{\rm p}$  & -0.357 &-0.2885 & -0.27897 & -0.2787  & -0.27839&-0.2784 &-0.2783 \\ \hline
                 $\mu _{\rm m}$  &  -7.42 &-7.39   & -7.387   & -7.385   & -7.39   & -7.39  &-7.39 \\ \hline
    \end{tabular}
    \caption{ }
    \label{table:2}
\end{table}

\bibliography{references}